\documentclass[aps,pre,onecolumn,groupedaddress,showpacs]{revtex4}
\usepackage{psfig}

\usepackage{psfig}

\begin{document}


\draft 
\title{Accuracy of Semiclassics: Comparative Analysis of WKB
and Instanton Approaches.}
 
\author{V.A.Benderskii} 
\affiliation {Institute of Problems of Chemical Physics, RAS \\ 142432 Moscow
Region, Chernogolovka, Russia} 
\affiliation{Laue-Langevin Institute, F-38042,
Grenoble, France} 
 
\author{E.V.Vetoshkin} 
\affiliation {Institute of Problems of Chemical Physics, RAS \\ 142432 Moscow
Region, Chernogolovka, Russia} 
\author{E. I. Kats} \affiliation{Laue-Langevin Institute, F-38042,
Grenoble, France} 
\affiliation{L. D. Landau Institute for Theoretical Physics, RAS, Moscow, Russia}
 
\date{\today}

\begin{abstract}
We analyze quantitatively the accuracy of eigenfunction and eigenvalue
calculations in the frame work of WKB and instanton semiclassical
methods. We show that to estimate the accuracy it is enough
to compare two linearly independent (with the same quantum number)
solutions to the Schr\"odinger equations with
the potential under study, and with the approximating
piecewise smooth potential. The main advantage of the approach
is related to the appropriate choice of the approximating potential,
providing absolutely convergent majorant series for the solutions. 
We test our method for a strongly anharmonic one dimensional potential, but the basic ideas inspiring our work and its results
can be applied to a large variety
of interesting chemical and physical problems
which are of 
relevance to various molecular systems.                                                                                        

\end{abstract}


\maketitle

\section{Introduction}
\label{int}

It is a textbook wisdom that if the de Broglie wavelengths $\lambda $ of particles
are small in comparison with the characteristic space scales of a given problem,
then the problem can be treated semiclassically. 
The commonly used WKB method (phase integral approach)
\cite{LL65} - \cite{HE62} is intended for the conditions of ''geometrical optics'',
in which the gradient of the action $\sigma $ is large, but slowly variable
(this is suggested also by it containing the factor $\hbar $, since we are dealing with a semiclassical
approximation, in which $\hbar $ is taken as small).
The corresponding condition
can be formulated more quantitatively \cite{LL65} as follows: $\lambda $ must vary only slightly
over distances of the order of itself
\begin{eqnarray}
\label{i1}
\left |\frac{d (\lambda (x)/2\pi )}{d x}\right | \ll 1
\, ,
\end{eqnarray}
where $\lambda (x) = 2\pi \hbar /p(x)$, and $p(x)$ is a classical momentum.
However, this simple criterion is not a practical tool to estimate
how accurate could be found semiclassical solutions of particular
problems, since nothing is specified regarding 
the convergence of the semiclassical series.
The criterion (\ref{i1}) does not work to estimate the magnitude
of the error involved in the approximate calculation of physical quantities (e.g., matrix elements),
neither to find the domain of validity in the complex plane in which
the semiclassical solutions are defined.
Indeed from (\ref{i1}) one can conclude only that
higher order corrections to semiclassical wave functions are small
in the asymptotic regions, but this mathematical criterion has almost
nothing to do with say the physical accuracy of semiclassical
matrix elements which depends on the wave function
accuracy in space regions providing main contributions
into the matrix elements under consideration.
For example, the energy eigenvalues are determined by the asymptotical 
region of the linear turning points (i.e., the region distant from these points),
and as well by the proximity region to the second order turning points, since in the
both regions the wave functions possess the largest values.

Within the WKB method such kind of a physical accuracy estimation has been performed
long ago by N. and P.O. Fr\"oman \cite{FF65}.
They analyzed higher order corrections to the semiclassical
wave functions and found that although those are really small over $1/\gamma ^2$
($\gamma \gg 1$ is semiclassical parameter), the
corrections are proportional to the factor 
$[(E /\gamma ) - U]^{-2}$,
(where $E $ is energy and $U$ is potential), and thus the function has non-integrable
singularity at the linear turning points where 
$[(E /\gamma ) - U] = 0$ (or, within the 
alternative to WKB semiclassical formalism so-called extreme tunneling trajectory or instanton
instanton approach \cite{PO77} - \cite{BV02},
the corrections are singular in the second order turning points). To surmount this problem in \cite{FF65}
(see also \cite{OL59}, and \cite{OL74}) the analytical continuation of the correction
function into the complex plane has been proposed, and it gives impractically
bulky expressions even for simple model potentials.
Since this problem has relevance far beyond WKB treatment of a particular
model potential, this is an issue of general interest to develop a simple and convenient
in practice quantitative method to study the accuracy of the semiclassical approach,
and it is the immediate motivation of the present paper 
to develop a systematic procedure how to do it.

The idea of our approach is to construct
two linearly independent continuous (with continuous first derivatives)
approximate solutions to the Schr\"odinger equation, which in the asymptotic
region coincide with semiclassic solutions, and in the vicinity of the
turning points - with the exact solutions of the so-called comparison
equation (i.e. the exact solution of the Schr\"odinger
equation for the chosen appropriately approximate near the turning
points potentials $V_c(X)$, henceforth will be referred to as the comparison potential). 
Although, scanning the literature we found one rather old paper \cite{LA37}
with a similar comparison equation approach, but our accuracy criterion is formulated as the majorant
inequalities for a certain matrix (which we find in the explicit analytical form
and calculated numerically) connecting our approximate 
and exact solutions in the finite space interval (not only in the vicinity of the turning points).
Since this method has largely gone unnoticed in the study
of semiclassics, we found it worthwhile to present
its derivation in a short and explicit form, and also to point out its practical usability.

The remainder of this paper is organized as follows.
In section \ref{theor} we present the basic
expressions necessary for our investigation. 
In this section we also present the main steps and qualitative idea of our method. 
Section \ref{calcul} contains our results. We derive the inequalities which enable us
to find the finite space interval (not at the isolated points) where the solutions have to be
matched, and calculate the $2 \times 2$ coordinate dependent matrix connecting
the approximate 
and exact solutions. 
Since the semiclassical solutions of the harmonic potential coincide with the exact
solutions, the accuracy of any semiclassically treated problem depends crucially on its
potential energy anharmonicity. That is why as the touchstone to test our method 
the results presented in
the section \ref{calcul} are applied to an anharmonic oscillator in section \ref{anh}.
We end with some brief
conclusions in the same section.

\section{Semiclassical equations in the WKB and instanton forms}
\label{theor}

Technically the basic idea 
how to overcome the difficulty of the semiclassical
solutions in the vicinity of the turning points
is reduced to 
an appropriate (admitting exact analytic solutions)
approximation of the potential
near the turning points.
After that step one has 
to match the asymptotics of this exact solution to the Schr\"odinger equation 
for an approximate potential with the
semiclassical solutions to the Schr\"odinger equation
for the potential under consideration (i.e. approximate solutions of the exact
potential) far from the turning points.
To illustrate main ideas of any semiclassic method (and to retain
compactness and transparency of expressions) we discuss
here a one dimensional case.
As it is well known \cite{LL65} in the WKB method solutions
to the Schr\"odinger equation are sought in the form
\begin{eqnarray}
\label{b1}
\psi = A \exp \left (\frac{i \sigma }{\hbar }\right )
\, ,
\end{eqnarray}
where for the function $\sigma $ called action the one particle 
Schr\"odinger equation (traditionally termed as Hamilton Jacoby 
equation)
reads as
\begin{eqnarray}
\label{b2}
\frac{1}{2m}\left (\frac{\partial \sigma }{\partial x}\right )^2
= E - U 
\, ,
\end{eqnarray}
where $m$ is a particle mass, $E$ is its energy, and $U$ is external
field potential.
Since the system is supposed quasi-classical in its properties, 
we seek $\sigma $ in the form of a series expanded in powers of
$\hbar $. 
Depending on normalization prefactor $A(x)$ entering (\ref{b1}) 
can be also found  
but the corresponding equation
(referred traditionally as transport equation) plays a pure
passive role since it is fully determined by the action $\sigma $
found as the solution of the Hamilton - Jacoby
equation
\begin{eqnarray}
\label{b22}
- \frac{i \hbar }{m}\left [\frac{1}{2}\frac{\partial ^2\sigma }{\partial x^2} A
+ \frac{\partial A}{\partial x}\frac{\partial \sigma }{\partial x}\right ]
+ \frac{\hbar ^2}{2 m}\frac{\partial ^2 A}{\partial x^2} = 0 
\, ,
\end{eqnarray}
where in the spirit of the semiclassical approximation the last term ($\propto \hbar ^2$)
is neglected.
Technically of course more convenient to use instead of $\hbar $
an expansion over equivalent but dimensionless parameter $\gamma ^{-1}
\ll 1$ we will call 
in what follows as semiclassical parameter and define as
\begin{eqnarray}
\label{b3}
\gamma \equiv \frac{m \Omega _0 a_0^2}{\hbar } \gg 1 
\, ,
\end{eqnarray}
where $a_0$ is a characteristic length of the problem, e.g. the tunneling distance,
$\Omega _0$ is a characteristic frequency, e.g. the oscillation frequency around the potential minimum.
Evidently the semiclassical parameter
$\gamma \gg 1$, and by its physical meaning it is 
determined by the ratio of the characteristic
potential scale over the zero oscillation energy.
We put $\hbar = 1$, and use $\Omega _0$ and $a_0$ to set corresponding dimensionless
scales, i.e. we introduce dimensionless energy $\epsilon \equiv E/\gamma \Omega _0$,
dimensionless coordinate $X \equiv x/a_0$, dimensionless potential $V \equiv U/\Omega _0$ (except
where
explicitely stated to the contrary and dimensions are necessary for
understanding or numerical estimations).

The analogous to (\ref{b1}), (\ref{b2}) procedure for the Schr\"odinger equation
in the imaginary time (instanton formalism, corresponding to the Wick rotation in the phase
space, when coordinates remain real valued $ x \to x$ but conjugated momenta
become imaginary $p_x \to i p_x$) can be
formulated as
the following substitution for the wave function (cf. to (\ref{b1}))
\begin{eqnarray}
\label{b4}
\psi = A_E(X) \exp (-\gamma \sigma _E)
\, ,
\end{eqnarray}
where the action $\sigma _E$ and we use the subscript $E$
to denote so-called Euclidean action obtained from the WKB action $\sigma $
after the Wick rotation.
Performed above rotation is not a harmless change of variables.
The deep meaning of this transformation
within the instanton approach is related to redistribution
of different terms between the Hamilton - Jacoby and the transport
equations. Indeed, like that is in the WKB method,
eigenvalues for the ground and for the low-lying
states are of the order of $\gamma ^0$, while all other terms
in (\ref{b2}) are of the order of $\gamma ^1$. Therefore to
perform a regular expansion over $\gamma ^{-1}$ for the substitution
(\ref{b4}) one has to remove the energy term from the Hamilton - Jacoby
equation, and to include this term into the transport equation.
Besides in the first order over $\gamma ^{-1}$
one can neglect the term with the second derivative of the prefactor.
As a result of this redistributions the both equations are
presented as
\begin{eqnarray}
\label{b5}
\frac{1}{2}\left (\frac{\partial \sigma _E}{\partial X}\right )^2 = V(X)
\, ,
\end{eqnarray}
instead of the WKB Hamilton - Jacoby equation (\ref{b2}),
and the transport equation is
\begin{eqnarray}
\label{b6}
\frac{\partial A_E}{\partial X}\frac{\partial \sigma _E }{\partial X}
+ \frac{1}{2}\frac{\partial ^2 \sigma _E}{\partial X^2} A_E = \epsilon A_E 
\, .
\end{eqnarray}
One can easily note by a simple inspection of the WKB (\ref{b2}), (\ref{b22})
and of the instanton (\ref{b5}), (\ref{b6}) equations that although
the both semiclassical methods can be formulated neglecting terms of the order
of $\gamma ^{-1}$, therefore possessing the same accuracy over $\gamma ^{-1}$,
the solutions evidently coincide in the asymptotic classically
forbidden region $ V(X) \gg \epsilon /\gamma $, but their behavior, number
and type of turning points (where any semiclassic approximation
does not work) are quite different.
For example in the WKB formalism there are  
two turning points where $V(X) - (\epsilon /\gamma ) = 0$ around each
minimum of the potential, while in the instanton
approach since the energy does not enter the Hamilton - Jacoby
equation (thus one can say that no classically accessible regions at all)
the turning points are extremal points (minima for the case) of the potential.
Furthermore as a consequence of this difference, in the WKB method
all turning points are linear, whereas in the instanton approach they are
second order (quadratic over $X$).

\section{Accuracy of semiclassical approximation}
\label{calcul}

Armed with this knowledge we are in the position now to construct our approximants.
Let us introduce besides the comparison potential
$V_c(X)$, one more specially chosen potential $V_{sc}$ (henceforth will be referred to as the semiclassical potential).
This potential is chosen by the requirement that the exact solutions to the Schr\"odinger
equation with $V_{sc}$ coincide asymptotically with the semiclassical solutions to the Schr\"odinger
equation with the potential $V(X)$ the problem under study.
Thus according to the construction, the semiclassical wave function $\Psi _{sc}$
satisfies the equation
\begin{eqnarray}
\label{ac1}
\Psi _{sc}^{-1}\frac{d^2 \Psi _{sc}}{d X^2} =
2 \gamma ^2 \left ( V_{sc}(X) - \frac{\epsilon }{\gamma }\right )
\, ,
\end{eqnarray}
and from here we can relate the semiclassical potential ($V_{sc}(X)$) with the bare one ($V(X)$)
\begin{eqnarray}
\label{ac2}
V_{sc}^{(1 , 2)} = V(X) \mp \frac{1}{2\gamma ^2} A^{-1}\left (\frac{d^2 A}{d X^2}\right )
\, ,
\end{eqnarray}
in the vicinity of the first order or of the second order (superscripts 1 or 2)
turning points.

Since near the turning point $X_0$ the prefactors $A^{(1)} \propto |X - X_0|^{-1/4}$, and
$A^{(2)} \propto (X - X_0)^n$ 
(where $n$ is an integer number which occurs from the transport equation (\ref{b6}) solution at the
energy $\epsilon = n + (1/2)$)
the potential $V_{sc}^{(1)}$ at $ X \to X_0$ is singular
and negative, and $V_{sc}^{(2)}$ has the same singularity ($\propto (X - X_0)^{-2}$) but positive.
The difference is due to the fact that near the WKB linear turning points we have deal
with the $V_{sc}^{(1)}$ well, whereas near the second order instanton turning points
one has to treat the potential barrier $V_{sc}^{(2)}$.
It might be useful to illustrate the essential features of the introduced
above potentials $V_c$ and $V_{sc}$ applying the definition (\ref{ac2}) to a simple
(but the generic touchstone) example of the following anharmonic oscillator
\begin{eqnarray}
\label{ac15}
V(X) = \frac{1}{2}[X^2 + \alpha X^3 + \beta X^4]
\, .
\end{eqnarray}
We show in Fig. 1 the semiclassical and the comparison potentials associated with (\ref{ac15})
for the WKB (Fig. 1a) and instanton (Fig. 1b) methods ($\alpha = -1.25$, $\beta = 0.5$,
and the energy window corresponds to $n=3$ excited state of the potential (\ref{ac15})).

The key elements to construct our approximants are the following
combinations related to probability flows to and from the turning points
\begin{eqnarray}
\label{ac3}
J(X_1) = \Psi _{sc}^{-1}\left (\frac{d \Psi _{sc}}{d X}
- \Psi _c^{-1}\frac{d \Psi _c}{d X}\right )_{X=X_1} = 2\gamma ^2 \Psi _{sc}^{-1}(X_1) \Psi _c^{-1}(X_1)
\int _{-\infty }^{X_1} \Psi _{sc}(X)\Psi _c(X) (V_{sc}(X) - V_c(X)) dX
\, ,
\end{eqnarray}
where $X_1 < X_0$ and analogously for $X_2 > X_0$ the flow function $J(X_2)$ is given by (\ref{ac3})
where the integration limits are from $X_2$ to $+ \infty $.
Since the exact wave functions are continuous with continuous first derivatives
(providing due to these features the continuity of the density probability currents),
the idea of our procedure is to require the same from the approximate wave functions.

The integrals entering $J(X_1)$ and $J(X_2)$ can be calculated easily for
any form of the potential, and the maximum accuracy of the any semiclassical approach
can be achieved upon the matching of the approximate solutions
at the characteristic points 
$X_{1 , 2}^\# $ 
where $J(X_{1 , 2}^\# ) = 0$.
The points $X_{1 , 2}^\#$ do exist in the case when the potentials $V_c$ and
$V_{sc}$ intersect in the region where the approximate wave functions $\Psi _{sc}$ and $\Psi _c$
are monotone ones. It is easy to realize (see e.g., Fig. 1)
that the both points occur 
in the vicinity of the linear turning points
for the potentials with $d^2V/d X^2 > 0$.
One such a point 
disappears
when the potential turning point becomes the inflection point, and there are no $X_{1 , 2}^\#$ points 
at all for
$d^2 V/d X^2 < 0$. 
In the vicinity of the second order turning point the comparison potential $V_c$ is a parabolic one.
The curvature of the latter potential
can be always chosen to guarantee the two intersection points
always exist. 
The choice of the comparison potential corresponds to a certain
renormalization ($\propto \gamma ^{-2}$) of the characteristic oscillation frequency
$(d^2 V/d X^2)_{X = X_0}$.
Note in passing
that the approach we are advocating here conceptually close (although not identical)
to the scale transformation proposed by Miller and Good \cite{MG53} and further
developed in\cite{PE71}.

Thus the conditions $J(X_{1 , 2}^\# ) = 0$ allow us to construct well controlled approximate
solutions to the Schr\"odinger equation. The accuracy of the approximation depends on the deviation
of the approximate wave functions from the exact ones in the vicinity of the characteristic points $X_{1 , 2}^\#$.
In own turn, the deviation is determined by the higher over $(X - X_0)$ terms of the potential $V(X)$ which are
not included in the harmonic comparison potential $V_c$.
Include explicitely the corresponding higher order terms to distinguish
$V_c$ and $V_{sc}$ potentials, we find in the vicinity of the linear turning points
\begin{eqnarray}
\label{ac4}
V_{sc}^{(1)} - V_c \simeq - \frac{c_1}{\gamma ^2}\left (X - X_0\right )^{-2} +
\frac{\omega ^2}{2} \left (X - X_0\right )^2
\, ,
\end{eqnarray}
where the universal numerical constant $c_1 = 5/32$, and the second term in the r.h.s.
is related to deviation of the bare potential from the linear one.
The same manner near the second order turning points
\begin{eqnarray}
\label{ac5}
V_{sc}^{(2)} - V_c \simeq - \frac{c_2}{\gamma ^2}\left (X - X_0\right )^{-2} +
\alpha (X - X_0)^3
\, ,
\end{eqnarray}
where the universal constant $c_2 = n(n-1)/4$ is zero for the lowest vibrational states
$ n = 0 \, , \, 1$, and the last term describes non-parabolicity of the potential.
Note that unlike the semiclassical action which within the instanton method is independent
of quantum numbers $n$, the position of the characteristic points does depend on $n$,
and the $X_{1 , 2}^\#$ points are placed near the boundaries of the classically accessible region.

Now we are in the position to construct the approximant wave functions
\begin{eqnarray}
\label{ac6}
\tilde {\Psi }(X) =
\left \{
\begin{array}{c}
\Psi _c(X) \, , \, X_1^\# < X < X_2^\# \\
\Psi _{sc} \, , \, X < X_1^\# \, , \, X > X_2^\#
\end{array}
\right .
\, ,
\end{eqnarray}
which are the solutions to the Schr\"odinger equation with the following
piecewise smooth approximating potential
\begin{eqnarray}
\label{ac7}
\tilde {V}(X) =
\left \{
\begin{array}{c}
V_c(X) \, , \, X_1^\# < X < X_2^\# \\
V_{sc} \, , \, X < X_1^\# \, , \, X > X_2^\#
\end{array}
\right .
\, .
\end{eqnarray}
The wave functions calculated according to (\ref{ac6}) in the framework
of the instanton approach close to the Weber functions in the classically
accessible regions, but their exponentially decaying tails
in the classically forbidden regions correspond to the exact (bare) potential, not to its
harmonic approximant. Analogously in the WKB method these functions (\ref{ac6}) coincide
with the semiclassical ones out of the interval $(X_1^\# \, , \, X_2^\# )$, and with the Airy
functions in this interval.

To proceed further on we have to relate our approximant wave functions
(\ref{ac6}) and two linearly independent solutions to the bare Schr\"odinger
equation $\Psi _1$ and $\Psi _2$. It can be written down formally as
\begin{eqnarray}
\label{ac8}
\Psi (X) = \tilde {\Psi }(X) + \int _{X_0}^{X} d X^\prime v(X^\prime ) G(X , X^\prime )
\Psi (X^\prime )
\, ,
\end{eqnarray}
where $v = V(X) - \tilde {V}(X)$,  $G(X , X^\prime )$ is the Green function for the Schr\"odinger
equation with the potential (\ref{ac7}), 
\begin{eqnarray}
\label{ins2}
G(X , X_1) \equiv const [\tilde {\Psi }_1(X)\tilde {\Psi }_2(X_1) - 
\tilde {\Psi }_1(X_1)\tilde {\Psi }_2(X)] \, ; \, X_1 \leq X  
\, ,
\end{eqnarray}
where the constant in (\ref{ins2}) is the Wronskian equal to $(2\gamma )^{-1}$ in the instanton method,
and $i(2\gamma )^{-1}$ within the WKB approach.
In (\ref{ac8}) $\Psi \equiv (\Psi _1 \, , \, \Psi _2)$
(the same definition for $\tilde \Psi $), and we take the turning point $X_0$, where
the functions $\Psi $ and $\tilde \Psi $ are close to each other at the lower integration limit.

The solution to the integral equation (\ref{ac8}) is expressed as the Neumann series
expansion, 
\begin{eqnarray}
\label{ins3}
\Psi (X) = \tilde {\Psi }(X) + \int _{X_0}^{X} dX_1 v(X_1) G(X , X_1)\tilde {\Psi }(X_1) + \cdots 
\, ,
\end{eqnarray}
and the $m$-th order term can be factorized and estimated as
\begin{eqnarray}
\label{ins4}
\leq \frac{1}{m!} \left (\int _{X_0}^{X} dX_1 v(X_1) G(X , X_1)\tilde {\Psi }(X_1)\right )^m
\, .
\end{eqnarray}
The integrals entering this estimation
\begin{eqnarray}
\label{ins5}
\int _{X_0}^{X} dX_1 v(X_1) G(X , X_1)\tilde {\Psi }(X_1) = L_{12/22}(X) \tilde {\Psi }_1(X) 
+  L{11/21}(X)\tilde {\Psi }_2(X) 
\, 
\end{eqnarray}
contain the $2 \times 2$ matrix with the following matrix elements
\begin{eqnarray}
\label{ac9}
L_{i j} = \int _{X_0}^{X} d X^\prime \tilde {\Psi }_i(X^\prime )v(X^\prime ) \tilde {\Psi }_j(X^\prime )
\, .
\end{eqnarray}
It is convenient to introduce the matrix $\hat C^{(n)}$ relating the $n$-th order 
wave function correction $\delta \Psi ^{(n)}$ with the wave function $\tilde \Psi $
\begin{eqnarray} &&
\label{ins6}
\left (
\begin{array}{c}
\delta \Psi _1^{(n)} \\
\delta \Psi _2^{(n)}
\end{array}
\right ) 
= \hat C^{(n)}
\left (
\begin{array}{c}
\tilde \Psi _1 \\
\tilde \Psi _2
\end{array}
\right ) 
\, ,
\end{eqnarray}
and the full connection matrix between the exact and approximate wave functions
\begin{eqnarray} &&
\label{ins7}
\left (
\begin{array}{c}
\Psi _1 \\
\Psi _2
\end{array}
\right ) 
= \hat C
\left (
\begin{array}{c}
\tilde \Psi _1 \\
\tilde \Psi _2
\end{array}
\right ) 
\, ,
\end{eqnarray}
is
\begin{eqnarray}
\label{ins8}
\hat C = \sum _{n} \hat C^{(n)}
\, .
\end{eqnarray}
According to the inequality (\ref{ins4})
\begin{eqnarray}
\label{ins9}
\hat {C}^{(n)} \leq \frac{1}{n!}\hat {C}_0^n
\, ,
\end{eqnarray}
where
\begin{eqnarray} &&
\label{ins10}
\hat C_0 =
\left (
\begin{array}{cc}
L_{12} & L_{11} \\
L_{22} & L_{12}
\end{array}
\right ) 
\, .
\end{eqnarray}
Combining finally the expressions (\ref{ins8}) - (\ref{ins10})
we end up with the upper and lower bounds for the correction matrix $\hat C$
estimation
\begin{eqnarray}
\label{ins11}
1 + \hat C_0 \leq \hat C \leq \exp (\hat {C}_0)
\, .
\end{eqnarray}
We conclude from (\ref{ins4}), (\ref{ac9}) that
the Neumann series posses an absolute convergence if all the matrix elements $L_{i j}$ are
finite. Besides, unlike the correction functions introduced within the Fr\"{o}man approach 
\cite{FF65}, the integrals in (\ref{ac9}) have no singularities on the real axis.
Evidently the integrals (\ref{ac9}) are finite with the oscillating WKB functions,
since the perturbation potential $v(X)$ is not zero only in the close proximity to
the characteristic points $X_{1 , 2}^\#$.
However in the
instanton method due to mixing of increasing and decreasing exponents,
the matrix elements 
$L_{22}$ is divergent.
Despite of this divergency the product $L_{22}\tilde {\Psi} _1$, we are only interested in,
is finite, and it is convenient to perform one more transformation
to exclude explicitely this divergency. 

Technically one can easily eliminate
the both off-diagonal elements of the matrix $\hat C_0$ and thus to get rid of the divergency of the 
(exponentially decreasing solution
$\tilde \Psi _1$) amplitude due to the contribution to the $\tilde \Psi _1$
the exponentially increasing solution $\tilde \Psi _2$.
These linear transformations renormalize the correction matrix elements $L_{22}$
and $L_{11}$ as follows
\begin{eqnarray}
\label{ac10}
L_{22}(X) \tilde {\Psi }_1(X)
\equiv L_{22}^*(X)
\tilde {\Psi }_2(X) 
\, ; \,
L_{11}(X) \tilde {\Psi }_2(X)
\equiv L_{11}^*(X)
\tilde {\Psi }_1(X) 
\, ,
\end{eqnarray}
where the renormalized matrix elements $L_{22}^*$ and $L_{11}^*$ read as
\begin{eqnarray}
\label{ac12}
L_{22}^*(X) = \int _{X_0}^{X} d X^\prime \frac{\tilde {\Psi }_1(X) \tilde {\Psi }_2(X^\prime )}{
\tilde {\Psi }_1(X^\prime ) \tilde {\Psi }_2(X)}
\tilde {\Psi }_1(X^\prime )v(X^\prime )\tilde {\Psi }_2(X^\prime ) \leq L_{12}(X)
\, ,
\end{eqnarray}
and
\begin{eqnarray}
\label{ac11}
L_{11}^*(X) = \int _{X_0}^{X} d X^\prime \frac{\tilde {\Psi }_1(X^\prime ) \tilde {\Psi }_2(X)}{
\tilde {\Psi }_1(X) \tilde {\Psi }_2(X^\prime )}
\tilde {\Psi }_1(X^\prime )v(X^\prime )\tilde {\Psi }_2(X^\prime ) \geq L_{12}(X)
\, .
\end{eqnarray}
Now all the integrals entering $L_{11}^*$, $L_{22}^*$, and $L_{12}$ are convergent and finite
at any $X$, 
the correction matrix $\hat C_0$ is transformed into the diagonal and positively defined
matrix $\hat {C}^*_0$
\begin{eqnarray} &&
\label{ins12}
\hat {C}^*_0 =
\left (
\begin{array}{cc}
L_{11}^* - L_{12} & 0 \\
0 & L_{12} - L_{22}^*
\end{array}
\right ) 
\, .
\end{eqnarray}
The $n$-th order corrections in the instanton approach satisfy the inequality
\begin{eqnarray}
\label{ac13}
0 \leq \delta \Psi _1^{(n)} \leq (n!)^{-1}(L_{11} - L_{12})\tilde {\Psi }_1
\, , \, 
0 \leq \delta \Psi _2^{(n)} \leq (n!)^{-1}(L_{12} - L_{22}^*)\tilde {\Psi }_2
\, .
\end{eqnarray}
Explicit summation of r.h.s in (\ref{ac13}) gives us the upper and the lower
bound limits for the solutions of the initial Schr\"odinger equation, i.e. the
stripe where increasing and decreasing solutions are confined
\begin{eqnarray}
\label{ac14}
|\tilde {\Psi }_1(X)|  \leq |\Psi _1(X)| \leq |\tilde {\Psi }_1(X)|\exp (L_{11}^* - L_{12})
\, , \, 
|\tilde {\Psi }_2(X)|  \leq |\Psi _2(X)| \leq |\tilde {\Psi }_2(X)|\exp (L_{12} - L_{22}^*)
\, .
\end{eqnarray}
It is our main result in this paper, and the stripe (\ref{ac14}) gives
the accuracy of the semiclassical instanton method.
Besides we are in the position now to estimate the 
contribution of increasing semiclassical solutions into decreasing ones (what is relevant
to solve eigenvalue problems). The summation convergent majorant series enables us to estimate
the upper bound for this contribution
\begin{eqnarray}
\label{ac141}
\frac{L_{11}^*}{L_{22}^*}\left (1- \exp (-L_{12})\right )
\, .
\end{eqnarray}
Therefore at $L_{12} \ll 1$ the summation of all order perturbation terms enhances
the 1-st order correction by the factor $L_{12}/L_{22}^*$.
Analogously the majorant estimates described above can be used to construct
the connection matrices linking the semiclassical solutions through the turning
points. The comparison of the bare connection matrices (see e.g.,
\cite{HE62}, \cite{BM94}, \cite{BV02}) with the matrices calculated accordingly to
(\ref{ac9}) - (\ref{ac12}) provides the estimates for the eigenvalue accuracy.

As it was mentioned already, for the WKB method
the procedure is even more simple, since no any divergency and therefore no need to perform the transformation
(\ref{ac10}). The similar to (\ref{ins4}) - (\ref{ins5}) factorization gives the $\hat C_0$ matrix
(cf. with (\ref{ins10}) for the instanton approach)
\begin{eqnarray} &&
\label{ins13}
\hat C_0 = i
\left (
\begin{array}{cc}
L_{12} & -L_{11} \\
L_{22} & -L_{12}
\end{array}
\right ) 
\, ,
\end{eqnarray}
and the estimations for the $n$-th order contribution (cf. with (\ref{ac13})
can be formulated now as
\begin{eqnarray} &&
\label{ins14}
|\hat C^{(2n)}| \leq \frac{1}{(2n)!}
\left (
\begin{array}{cc}
\Delta & 0 \\
0 & \Delta 
\end{array}
\right )^n 
\, ,
\end{eqnarray}
and
\begin{eqnarray} &&
\label{ins15}
|\hat C^{(2n+1)}| \leq \frac{1}{(2n+1)!}
\Delta ^n \hat C_0
\, ,
\end{eqnarray}
where we denote $\Delta \equiv L_{12}^2 - L_{11}L_{22}$.
Correspondingly to (\ref{ins14}), (\ref{ins15}) the diagonal and off-diagonal 
correction matrix
elements are bounded from above
\begin{eqnarray} &&
\label{ins16}
|C_{11/22}| \leq \left |\cos \sqrt \Delta + i L_{12} \frac{\sin \sqrt \Delta }{\sqrt \Delta }\right |
\, ,
\end{eqnarray}
and
\begin{eqnarray} &&
\label{ins17}
|C_{12/21}| \leq \left |L_{11} \frac{\sin \sqrt \Delta }{\sqrt \Delta }\right |
\, .
\end{eqnarray}
The whole procedure we employed is rationalized in the Fig. 2, where we compare
the solutions to the comparison equation with the anharmonic oscillator semiclassical wave functions
computed within the instanton and WKB approaches and indicate the optimal matching points $X^\# $
found accordingly to the condition $J(X^\# ) = 0$. 

\section{Anharmonic oscillator}
\label{anh}
In closing let us illustrate how our estimations (\ref{ac14}), (\ref{ins15}) work for a strongly anharmonic potential
(\ref{ac15}). 
Although it is not great triumph to re-derive the known results, our derivation illustrates several characteristic features 
of the correction
matrix techniques derived in the section \ref{calcul}: better accuracy, rapid convergence, simple disposal of divergences, and ease
of computation in particular.
The main message of our consideration in the precedent section \ref{calcul} is that the quantitative
accuracy of the semiclassics depends crucially on the proximity
of the semiclassical wave functions to the solutions of the comparison equation
in the region of the asymptotically smooth matching.
Therefore, it is tempting to improve the accuracy by taking into account
the anharmonic corrections to the comparison potential $V_c(X)$.
However, since the eigenvalues and the normalization of the wave functions are almost independent
of the detailed behavior in the vicinity of the linear turning points
(because near these points, situated at the boundaries of the classically accessible
region, the probability density (i.e. $|\Psi |^2$) is exponentially small) this idea is useless
for the WKB approach.
In contrast with this, for the instanton method, the accuracy can be improved considerably
upon including the anharmonic corrections into the comparison potential.
Indeed, within the instanton method the accuracy is determined by the vicinity
of the second order turning points where the wave functions acquire the largest values
(and just in this region the smooth matching described above has to be performed).

Let us remind first the traditional (but formulated within the semiclassical framework)
perturbation theory. Keeping in the transport equation (\ref{b22}) the second derivative 
of the prefactor $A$,
the system of the equations of Hamilton - Jacoby (\ref{b2}) and the exact
transport equation  
\begin{eqnarray}
\label{anh1}
\frac{d^2 A}{d X^2} - 2 \gamma \left [\frac{d A}{d X}\frac{d \sigma }{d X}
+ \left (\frac{d^2 \sigma }{d X^2} - \epsilon \right )A\right ] = 0 
\, 
\end{eqnarray}
are exactly equivalent to the Schr\"odinger equation under consideration.
For the second order turning points the anharmonic corrections 
\begin{eqnarray}
\label{anh2}
V_p = \frac{1}{2}\left (\alpha X^3 + \beta X^4\right ) 
\, 
\end{eqnarray}
can
be considered as a perturbation and it is convenient to include this perturbation
$V_p$ into the transport equation (\ref{anh1}). 
Then the comparison equation is reduced to the inhomogeneous Weber equation,
and its solutions can be expanded over the Weber functions $D_\nu (X)$ \cite{EM53}
(see also \cite{OL59}, \cite{OL74})
\begin{eqnarray}
\label{anh3}
\Psi _\nu (X) = N_\nu \left (D_\nu (X) + \sum _{k} b_{k \nu } D_{\nu + k}(X) \right ) 
\, ,
\end{eqnarray}
where $N_\nu ^{-2} = 1 + \sum _{k} b_{\nu k}^2$ is the wave function normalization
factor, the expansion coefficients are proportional to the small parameters $\alpha /\sqrt \gamma $,
and $\beta /\gamma $, and the Weber function index $\nu $ is related to the energy eigenvalue $\epsilon = \nu + (1/2)$.
This expansion (\ref{anh3}) looks like a conventional perturbation series, but it does not. In the comparison
equation we are keeping the both (decreasing and increasing) waves, and as a result of it, the indices
of the Weber functions $\nu + k$ are not integer numbers.
In the first order over the perturbation $V_p$ the only non-zero coefficients in (\ref{anh3})
correspond to the following selection rules
\begin{eqnarray}
\label{anh4}
k = \pm 1 , \pm 3 \, ; \, {\rm {and}}\,  k =0 , \pm 2 , \pm 4  
\, 
\end{eqnarray}
for the cubic and fourth order anharmonic corrections respectively.
Explicitely these non-zero expansion coefficients can be found by straitforward
calculations, and they are
\begin{eqnarray}
\label{anh5}
b_{0 \nu } = -\frac{3}{2}\beta \left (\nu ^2 + \nu + \frac{1}{2}\right )\, ,
\, b_{-1 \nu } = - 3 \alpha \nu ^2 \, , \, b_{1 \nu } = - \alpha (\nu + 1) \, , \,
b_{-2 \nu } - \frac{\beta }{2}(\nu - 1)\left (\nu - \frac{1}{2}\right ) \, , \,
b_{2 \nu } - \frac{\beta }{2}\left (\nu + \frac{3}{2}\right )\, , \,
\end{eqnarray}
\begin{eqnarray}
\nonumber
b_{-3 \nu } = - \frac{1}{3}\alpha \nu (\nu -1)(\nu - 2)(\nu - 3) \, , \,
b_{3 \nu } = - \frac{1}{3}\alpha \, , \,
b_{-4 \nu } = -\frac{1}{4}\beta \nu (\nu -1)(\nu -2)(\nu -3)(\nu -4) \, , \,
b_{4 \nu } = -\frac{1}{4}\beta  
\, .
\end{eqnarray}
On equal footing we can find the perturbative corrections
to the Bohr-Sommerfeld quantization rules, and therefore the eigenvalues.
The calculation is straightforward, though deserves some precaution and rather tedious.
Skipping a large amount of tedious algebra
we end up with the fractional part of the quantum number $\nu $
\begin{eqnarray}
\label{anh6}
\nu \equiv n + \chi _n  
\, ,
\end{eqnarray}
and up to the second order over the anharmonic perturbation $V_p$ we find
\begin{eqnarray}
\label{anh8}
\chi _n^{(2)} = - \frac{15 \alpha ^2}{2 \gamma }\left (n^2 + n + \frac{11}{30}\right )
+ \frac{3\beta }{\gamma }\left (n^2 + n + \frac{1}{2} \right )
\, . 
\end{eqnarray}
However the described standard perturbative approach leads
to qualitatively wrong features of the solutions.
For example, the wave functions (\ref{anh3}), (\ref{anh5}) are represented as
a product of $\nu $ independent exponential factors and dependent of $\nu $ polynomials.
As it is well known \cite{LL65} in one dimension the $n$-th excited state wave function
must have $n$ zeros (and the number of zeros may not be changed by any perturbation).
However in the $m$-th order perturbation theory approximation, the wave function (\ref{anh3})
corresponding to a certain excited state $n$
contains Hermitian polynomials up to the order $n + 3m$ or $n+4m$ for the cubic
or quartic anharmonic perturbations, respectively.
Therefore some false zeros of the wave function appears in the standard perturbation theory, and
the region where the function oscillates becomes more and more wide in the higher
order over perturbations approximation. The contributions of these qualitatively
and quantitatively incorrect higher order terms become dominating in the asymptotic region
at $|\alpha \gamma | \sim 1$, $\beta \gamma \sim 1$. It conforms with the classical results due to Bender and Wu
\cite{BW69}, \cite{BW73} who have shown that for the quartic anharmonic potential ($\alpha = 0$, $\beta > 0$ in (\ref{ac15}))
the convergency radius is zero. Moreover, for $\beta =0$ and for an arbitrary small $\alpha $ 
(\ref{ac15}) is the cubic
anharmonic potential, i.e., the decay one. Thus it should have only complex eigenvalues,
what is not the case for the eigenvalues calculated within the perturbation theory.
The method we developed in section \ref{calcul} enables us not only
to estimate more accurate
the anharmonic corrections to the eigenvalues, and to bring
the whole schema of the calculations in a more elegant form. 
Our finding of the correction matrices
is not merely to surpass a technical difficulty of the standard perturbative method, it is more one of principle,
and we will show that the method has no drawbacks of the perturbation theory.

The proof proceeds as follows. Let us consider first the instanton method for the anharmonic potential (\ref{anh1})
possessing one second order turning point $X =0$.
As it was shown in the section \ref{calcul} one has to find also two other characteristic points which are
the roots of the equation (\ref{ac3}). We denote the
points as $X_L^\# $, and $X_R^\# $ (to refer by the self-explanatory subscripts $L$ and $R$ to the
left and to the right from the turning point $X=0$).
At the next step using the correction matrices introduced in the section \ref{calcul}, we can define formally
the transformation of our approximate wave functions (\ref{ac6}) into the unknown exact
wave functions $\Psi $ as
\begin{eqnarray} &&
\label{anh16}
\left (
\begin{array}{c}
\Psi _L \\
\Psi _R
\end{array}
\right ) 
= \hat C
\left (
\begin{array}{c}
\tilde \Psi _L \\
\tilde \Psi _R
\end{array}
\right ) 
\, ,
\end{eqnarray}
where as above
the subscripts $L$ and $R$ refer to the wave functions in the regions to the
left and to the right from the turning point $X=0$.
We do not know the correction matrix $\hat C$ but we do know
(see (\ref{ins6}) - (\ref{ac11})) the boundary estimations
for the matrix.

The Eq. (\ref{anh16}) can be used also to correct the known at the second order
turning point the connection matrix $\hat M$ \cite{HE62}, \cite{BV02}.
Indeed the connection matrices link the semiclassical solutions
in the $X$-regions to the left and to the
right from the turning points. For the isolated second order turning point
(we are dealing within the instanton method), the connection matrix $\hat M$ links the exponentially
increasing and decreasing solutions in the space regions separated by the turning point.
The condition ensuring the correct asymptotic behavior is the quantization rule for this case
which can be formulated as $M_{11}=0$ ($M_{ij}$ are the matrix elements of the connection matrix $\hat M$).
Since in the regions to the left and to the 
right from the turning point our approximate solutions $\tilde \Psi $
coincide by their definition (\ref{ac6}) with the semiclassical ones, the correction matrix method
enables us to correct the quantization rule too. Namely, the quantization rules are formulated within the connection matrix technique
read now as
\begin{eqnarray}
\label{anh7}
C_{22}^R T_2 C_{22}^L - C_{21}^R C_{21}^L \frac{\sin ^2(\pi \nu )}{T_2} +
(C_{22}^L C_{21}^R + C_{21}^L C_{22}^R)\cos (\pi \nu ) = 0  
\, ,
\end{eqnarray}
where the Stokes constant for the second order turning point \cite{HE62} is 
\begin{eqnarray}
\label{anh77}
T_2 = \frac{\sqrt {2 \pi }}{\Gamma (-\nu )}) 
\, ,
\end{eqnarray}
and $C_{ij}^{R , L}$ are the correction matrices at
the $X_R^\# $ or $X_L^\# $ characteristic points respectively.
Expanding the Gamma function entering (\ref{anh7}) around the integer numbers, i.e., as above (\ref{anh6}), $\nu = n + \chi _n$
one can find from the equation (\ref{anh7}) the fractional part of the quantum number.
If we were known the correction matrix $\hat C$ the solution of (\ref{anh7}) would provide the exact
eigenvalues. But we do know only the estimations from below and from above for the $\hat C$ matrix.
In the same spirit we can calculate the estimations for the correction matrices
(and therefore for the eigenvalues) within the WKB approach.

Luckily 
it turns out that the mathematical nature of the semiclassical problem
is on our side here, and, in fact, even the first order estimation from below
$\hat C^{(0)}$ (\ref{ins6}) gives already the accuracy comparable with the standard
perturbation procedure, and the estimation from above (\ref{ins8}) gives the eigenvalues
almost indistinguishable from the ''exact'' ones obtained by the numerical
diagonalization of the Hamiltonian.
The same true for the wave functions found by the correction matrix technique.
We show in Fig. 3
$|\Psi _3|^2$ for the same anharmonic potential (\ref{ac15}) with $\alpha = -1.25$, $\beta = 0.5$.
Clearly the exact numerical results and those obtained by our correction matrix techniques
are correct qualitatively and in the very good quantitative agreement (indistinguishable
starting from the second order approximation) unlike the situation with the standard
perturbation theory.
Besides we present in the table the eigenvalues of the  anharmonic potential.
We take the anharmonic coefficients $\alpha $ and $\beta $ in (\ref{ac15}) so large, that
corresponding perturbations of the eigen values are of the order of 
the bare harmonic frequency (one in our dimensionless units $\alpha = - 1.2 \, , \, \beta = 0.5 $).
In the table the eigenvalues 
found by the numerical diagonalization are presented in the column $I$.
The column $II$ contains the harmonic approximation results, the column $III$ is
the second order perturbation theory (\ref{anh8}), and the columns $IV$ and $V$ results are obtained by applying our
correction matrix technique: estimation from below with the first order correction matrix (\ref{ins6}) in the
column $IV$,
and the estimation from above with the matrix (\ref{ins8}) in the column $V$.

To conclude, as we have shown how to estimate the corrections to the main technical
tool for the semiclassical approach, the connection matrices linking
the solutions to the left and to the right from the turning points.
Everything (e.g., the upper and the lower bounds for $\chi _n$)
is determined by the matrices $L_{12}$ and $L_{22}^*$.
Approximating to (\ref{ac15}) potential is found from (\ref{ac4}) and after
that straightforward computing according to (\ref{ac9}) the matrices $L_{ij}$ and their renormalization (\ref{ac12})
leads to the corrections we are looking for, presented in the Fig. 4, which allow us to estimate 
the accuracy of the semiclassical eigenstates and eigenfunctions.
We conclude that even for a strongly
anharmonic potential the both methods (WKB and instanton) are fairly accurate ones
(about $5\%$) up to the energy close to the potential barrier top (in the region
of negative curvature, we already discussed above). 
It is worth noting that in the frame work of the conventional perturbation theory
(due to zero convergency radius with respect to $\beta $ coefficient in (\ref{ac15}))
pure computational problems to get the same accuracy become nearly unsurmountable,
see e.g., \cite{BW69}, \cite{TU84}.
Our findings show that to estimate quantitatively
the semiclassical accuracy it is enough
to compare two linearly independent (with the same quantum number)
solutions of the initial potential under study, and of the approximating
piecewise smooth potential. The main advantage of the approach
is related to the appropriate (\ref{ac4}), (\ref{ac7}) choice of the approximating potential,
providing absolutely convergent majorant series (\ref{ac8}) for the solutions. 
Actually our correction matrix technique is a fairly universal one and enables
to estimate (and improve!) the semiclassical accuracy
for arbitrary one dimensional potentials with any combination
of the turning and of the crossing points.

\acknowledgements 
The research described in this publication was made possible in part by RFFR Grants. 
One of us (E.K.) is indebted to INTAS Grant (under No. 01-0105) for partial support,                            
and V.B. and E.V. are thankful to CRDF Grant RU-C1-2575-MO-04.

\newpage

\centerline{Figure Caption}

Fig. 1

The characteristic semiclassical potentials $V_{sc}$ (dot-dashed lines) and $V_c$ (dashed lines) for the bare anharmonic 
potential (\ref{ac15}): $\alpha = -1.25 $, $\beta = 0.5$ ($\gamma = 33$, and the energy window
corresponds to $n=3$):

(a) instanton approach;

(b) WKB method.

Fig. 2

Semiclassical wave function $\Psi _3$ ($n=3$) for the anharmonic potential
(\ref{ac15}) with $\alpha = - 1.25 $, $\beta = 0.5$ ($\gamma = 33$).
Stars indicate the matching points, dashed lines show the solutions to the comparison equations,
and:

(a) - solid line traces the instanton solution;

(b) - solid line shows the WKB wave function.

Fig. 3

Comparison of the exact $|\Psi _3|^2$ (solid line) 
for the anharmonic potential (\ref{ac15}) with $\alpha = -1.25$, $\beta = 0.5$ ($\gamma = 33$)  
with two lowest (zero and first order) approximations of the correction matrix method
(dashed and dot-dashed lines). Note that the second order approximation with the relative accuracy $10^{-2}$
is indistinguishable from the exact numerical results.

Fig. 4

Corrections to the decreasing solutions in
for the anharmonic potential (\ref{ac15});

$\beta = 0.5 \, , \, n =3 \, , \, \gamma = 33 $;

(1, 3) - instanton method, (2, 4) - WKB, 
(1, 2) - the first order corrections $L_{11}$, (3 , 4) - the upper bound estimation
summing up all terms. The vertical dashed lines indicate
the values of the cubic anharmonic term ($\alpha $) where the inflection point and
new extrema of the potential are appeared.

Table

Eigenvalues of the  anharmonic potential (\ref{ac15}) ($\alpha = - 1.2 \, , \, \beta = 0.5 $).

$I$ the eigenvalues 
found by the numerical diagonalization;

$II$ the harmonic oscillator eigenvalues;

$III$ the eigenvalues in the second order perturbation theory (\ref{anh8});

$IV$ the eigenvalues estimated from below by the correction matrix (\ref{ins6});

$V$ the estimation from above with the matrix (\ref{ins8}).


\begin{references}
\bibitem{LL65} L.D.Landau, E.M.Lifshits, Quantum Mechanics (non-relativistic
theory), Pergamon Press, New York, 1965. 
\bibitem{FH65} R.P.Feynman, A.R.Hibbs, Quantum Mechanics and Path Integrals,
McGraw-Hill Book Company, New York, 1965.
\bibitem{HE62} J.Heading, An Introduction to Phase-Integral
Methods, Wiley - Interscience, London, 1962.                                                                   
\bibitem{FF65} N.Fr\"oman, P.O.Fr\"oman, JWKB Approximation, North-Holland
Publishing Company, Amsterdam, 1965.
\bibitem{PO77} A.M.Polyakov, Nucl.Phys. B, {\bf 129}, 
429 (1977).
\bibitem{CO85} S.Coleman, Aspects of Symmetry, Cambridge University Press,
Cambridge, 1985.
\bibitem{SC86} A.Schmid, Ann. Phys., {\bf 170}, 333 (1986).
\bibitem{BM93} V.A.Benderskii, D.E.Makarov,
P.G.Grinevich, 
Chem. Phys., {\bf 170}, 275 (1993).
\bibitem{BM94} V.A.Benderskii, D.E.Makarov, C.A.Wight, Chemical
Dynamics at Low Temperatures, Willey-Interscience, New York, 1994.
\bibitem{BV00} V.A.Benderskii, E.V.Vetoshkin, 
Chem. Phys., {\bf 257}, 203 (2000). 
\bibitem{BV02} V.A.Benderskii, E.V. Vetoshkin, E.I.Kats, JETP, {\bf 95}, 645 (2002).
\bibitem{LA37} R.E.Langer, Phys. Rev., {\bf 51}, 669 (1937).
\bibitem{OL59} F.W.J.Olver, J. Res. Nat. Bur. Stand., {\bf 63 B}, 131 (1959).
\bibitem{OL74} F.W.J. Olver, Asymptotics and Special Functions,
Acad. Press., New York, 1974.
\bibitem{MG53} S.C.Miller, R.H.Good, Phys. Rev., {\bf 91}, 174 (1954).
\bibitem{PE71} P.Pechukas, J.Chem. Phys., {\bf 54}, 3864 (1971).
\bibitem{BW69} C.M.Bender, T.S.Wu, Phys. Rev., {\bf 184}, 1231 (1969).
\bibitem{BW73} C.M.Bender, T.S.Wu, Phys. Rev. D, {\bf 1}, 1620 (1973).
\bibitem{TU84} A.V.Turbiner, Sov. Phys. Usp., {\bf 27}, 668 (1984).
\bibitem{EM53} A.Erdelyi,
W.Magnus, F.Oberhettinger, F.G. Tricomi,
Higher Transcendental Functions, vol.1 - vol.3, McGraw Hill, New
York (1953).
\bibitem{BV04} V.A.Benderskii, E.V. Vetoshkin, E.I.Kats, Phys. Rev. A, {\bf 69}, 062508 (2004).


\end{references}
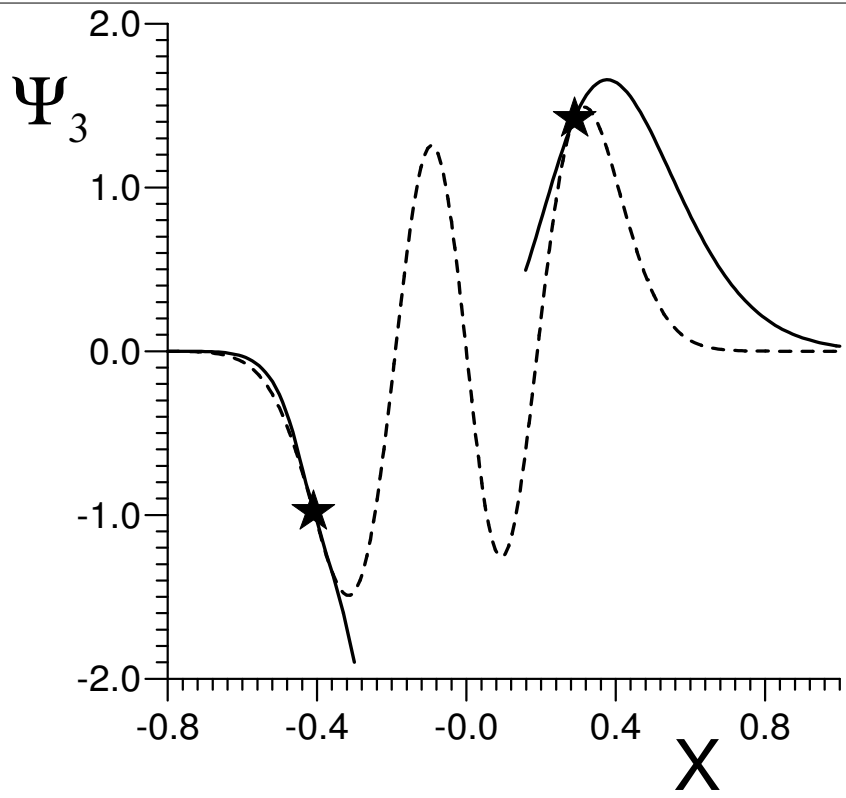
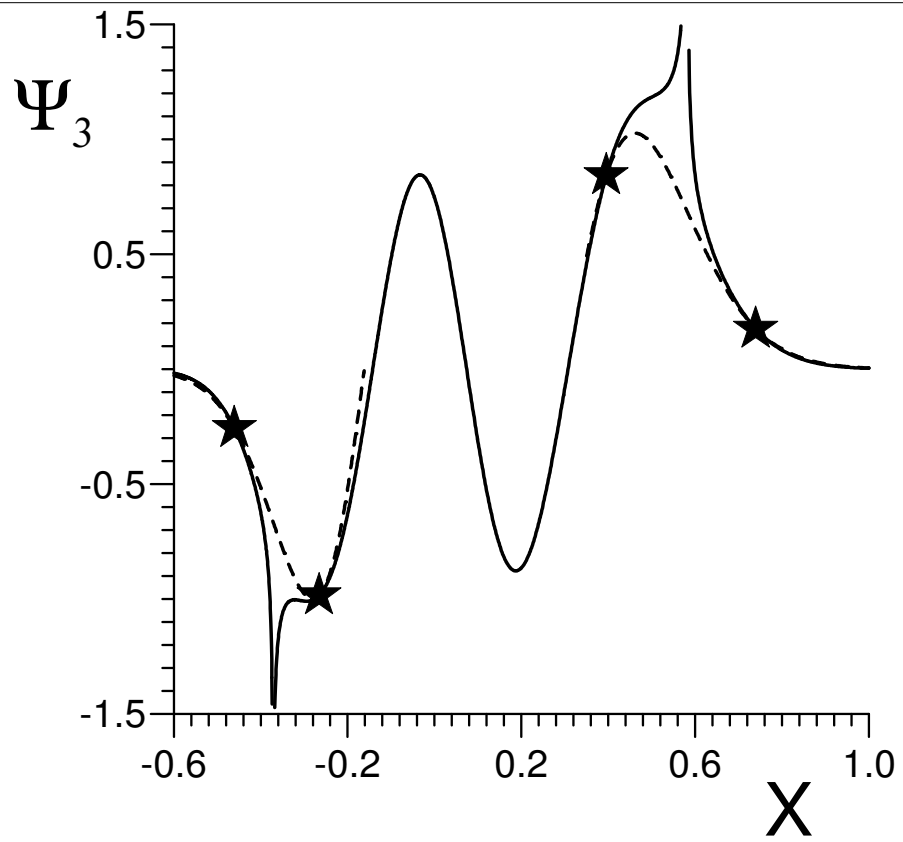
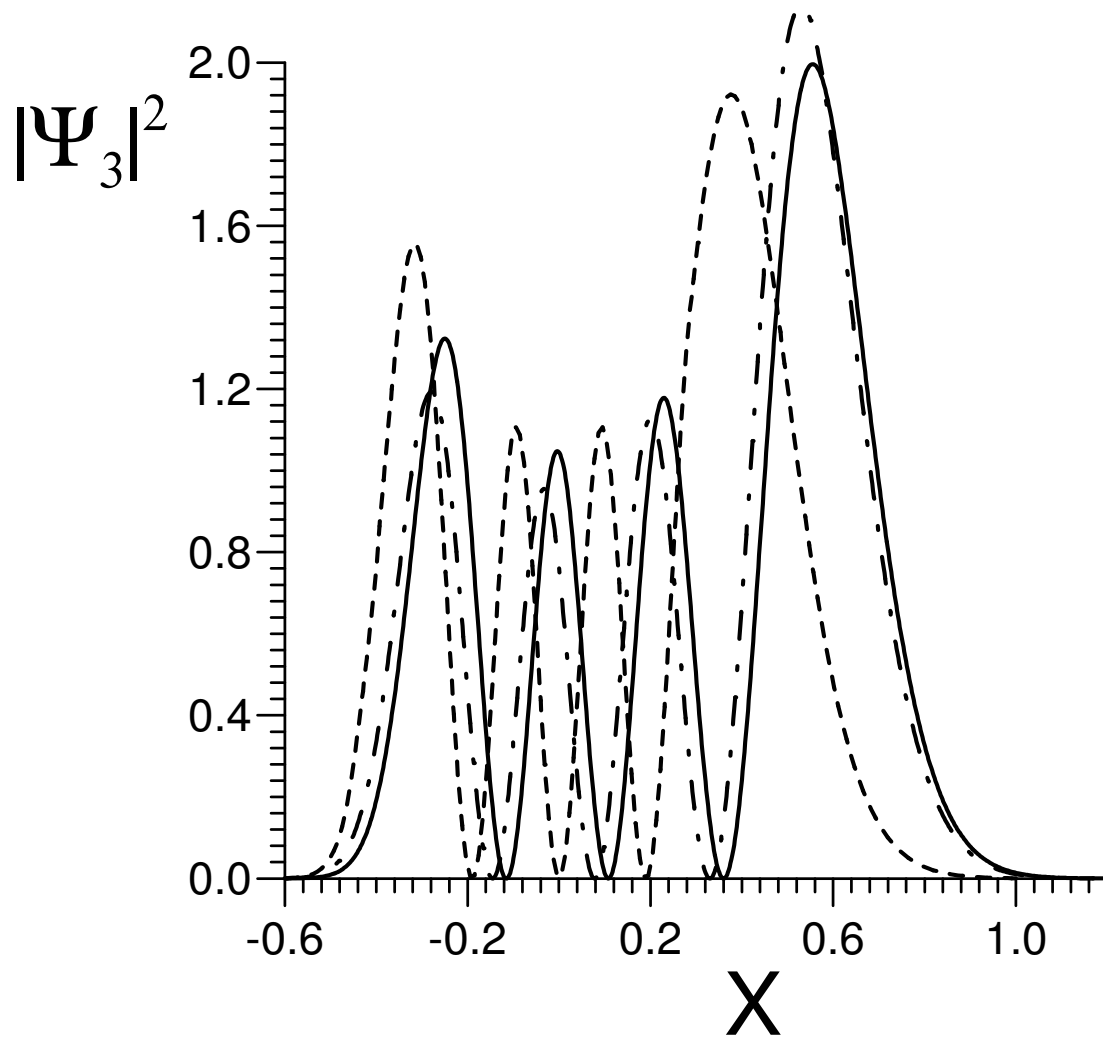
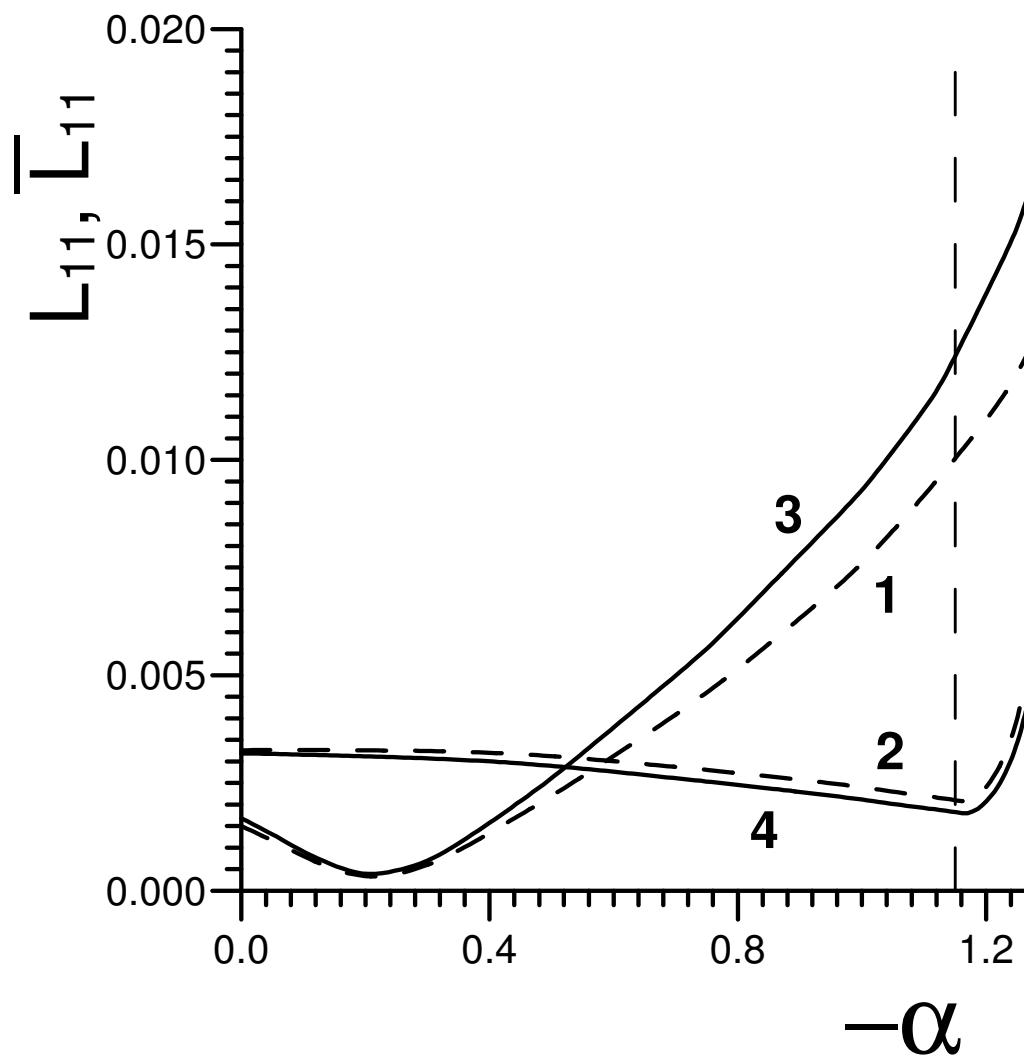
\end{document}